\documentclass[aps,prl,preprint]{revtex4}%
\usepackage{amsfonts}
\usepackage{amsmath}
\usepackage{amssymb}
\usepackage{graphicx}%
\begin{document}
\title[Spin Hall Effect ]{Spin Hall Effect induced by resonant scattering on impurities in metals}
\author{Albert Fert}
\affiliation{Unit\'{e} Mixte de Physique CNRS/Thales, 91767, Palaiseau, France and
Universit\'{e} Paris-Sud, 91405, Orsay, France}
\author{Peter M Levy}
\affiliation{Department of Physics, New York University, 4 Washington Place, New York, NY 10003}
\keywords{Hall effect, spin currents }
\pacs{PACS numbers: 85.75.-d,75.76.+j,73.50.Jt}

\begin{abstract}
The Spin Hall Effect (SHE) is a promising way for transforming charge currents
into spin currents in spintronic devices. Large values of the Spin Hall Angle,
the characteristic parameter of the yield of this transformation, have been
recently found in noble metals doped with nonmagnetic impurities. We show that
this can be explained by resonant scattering off impurity states split by the
spin-orbit interaction. We apply our calculation to the interpretation of
experiments on copper doped with 5d impurities and we describe the conditions
to obtain the largest effects.

\end{abstract}
\date[Date text]{date}
\startpage{1}
\endpage{2}
\maketitle
\accepted{}

The Spin Hall Effect (SHE), first described by Dyakonov and Perel in
1971\cite{1}, is a subject of intense research as it allows for the generation
of spin currents in nonmagnetic conductors and the developments of spintronic
devices built without ferromagnetic materials. The SHE is due to spin-orbit
(S-O) interactions which deflect the spin up and spin down electrons of an
electrical current in opposite directions. While the symmetry between spin up
and spin down in a nonmagnetic material precludes charge accumulation on the
edges of the conductor, i.e. no Hall voltage, there is a spin accumulation
which can be exploited to generate a pure spin current. Alternatively a Hall
voltage can be generated by injecting a spin-polarized current to break the
symmetry, which is called the Inverse SHE effect. The SHE is associated with
off-diagonal terms of the resistivity tensor having opposite signs for spin up
and spin down electrons, respectively $\rho_{xy}$ and -$\rho_{xy}$for $s_{z}$
= $\pm$1/2. It can include an intrinsic contribution due to the effect of S-O
interactions on the wave functions of the pure material \cite{2,3} and an
extrinsic one resulting from spin-orbit interactions on impurity or defect
sites\cite{4,5}. Two mechanisms can contribute to the extrinsic SHE, the skew
scattering \cite{4} and the scattering with side-jump \cite{5}.

When the SHE is used to produce a transverse spin current, the maximum yield
of the transformation of a longitudinal charge current into a transverse spin
current is related to the Spin Hall Angle (SHA), defined as $\Phi_{H}%
=\rho_{xy}/\rho_{xx}$ where $\rho_{xx}$ is the diagonal term of the
resistivity tensor, i.e., the conventional resistivity for spin $\sigma
=\uparrow\downarrow(\pm)$ channels. Consequently $\Phi_{H}$ is the important
parameter for practical applications in spintronics. Until 2007 the largest
values of $\Phi_{H}$ obtained for pure materials, metals or semiconductors,
had been obtained for Pt ($\Phi_{H}$ $\thickapprox0.5\%$~) \cite{6,7}. The
much larger value of $5\%$ found in 2008 for Au \cite{8} was surprising and
has been ascribed to skew scattering by Fe or Pt impurities \cite{9}. An even
larger SHE ($\approx15\%$) was recently obtained by doping Au with Pt
impurities \cite{10}. Actually this brings to mind the large values of
$\Phi_{H}$ of a few percent found thirty years ago~\cite{11,12} for the SHE
induced by nonmagnetic 5d impurities in Cu, e.g. $\Phi_{H}$ $=2.6\%$, for Cu
doped with Ir. This large SHE, with a typical change of sign between the
beginning (Lu) and the end (Ir) of the 5d series, was ascribed to resonant
scattering on the impurity 5d states split by the S-O interaction \cite{11}.
Recently measurements by Niimi et al \cite{13} on Cu doped with Ir have
confirmed the large value ($\Phi_{H}$ $\thickapprox1.5\%$) of the SHE induced
by Ir in Cu and confirmed its skew scattering mechanism. Thus, impurity
scattering appears as a promising way to obtain the most efficient
transformation of charge currents into spin currents by SHE. This has
triggered the development of theoretical models of the SHE induced by
impurities~\cite{14}.

In this paper we present a calculation and discussion of the SHE induced by
resonant scattering from impurity levels. For 5d impurities in Cu, we can
explain the order of magnitude of the large SHE of the experiments and the
change of sign between the beginning and the end of the 5d series. Whereas the
recent papers of Gradhand et al. \cite{14} present calculations of only the
skew scattering contribution, we calculate both the skew scattering and side
jump terms. By comparing the spin Hall angles due to skew scattering and to
side-jump, we can predict the threshold concentration at which the side-jump
contribution becomes predominant and can generate very large effects. In
contrast to the ab-initio calculations of Ref.~\cite{14}, our calculation is
performed in an analytical model which aims at a general description of the
main features of the impurity-induced SHE and at a prediction of the best
conditions for large effects.

Our calculation is based on a partial wave analysis of the resonant scattering
of free electrons from the $j=5/2$ and $j=3/2$ states of $5d$ impurities in a
metal like Cu, as illustrated in the inset of Fig.1. From the splitting
between the $5/2$ and $3/2$ levels, $E_{5/2}-E_{3/2}=5\lambda_{d}/2$, where
$\lambda_{d}$ is the impurity $5d$ S-O constant and by using the classical
expression of the phase shift at energy $E$ as a function of the resonant
level energy $E_{j}$, $ctn(\eta_{j})=(E_{j}-E)/\Delta$ where $\Delta$ is the
resonance width, we find to first order in $\lambda d/\Delta$ , $\Delta
\eta=\eta_{3/2}-\eta_{5/2}=5/2\frac{\lambda_{d}}{\Delta}\sin^{2}\eta_{2}$
where $\eta_{2}$ is the mean phase shift expressed as a function of the
number$Z_{d}$ of $5d$ electrons on the impurity by Friedel's sum rule,
$\eta_{2}=$ $(3\eta_{3/2}+2\eta_{5/2})/5=\pi Z_{d}/10$ \cite{15}. After
expanding the states $\left\vert j,m_{j}\right\rangle $ in terms of
$\left\vert m,\sigma\right\rangle $ states and keeping only terms that will
contribute to $\rho_{xx}$ and $\rho_{xy}$ ,we find the following expression
for the scattering $T$-matrix (to first order in $\lambda_{d}/\Delta$ only
non-spin-flip terms contribute),%

\begin{equation}
T_{\mathbf{k}^{\prime}\sigma,\mathbf{k}\sigma}=\frac{2}{n(\varepsilon_{F}%
)}\left[  \sigma\frac{\lambda_{d}}{\Delta}e^{i2\eta_{2}}\sin^{2}\eta_{2}%
\sum_{m}mY_{2}^{m\ast}(\hat{k})Y_{2}^{m}(\hat{k}^{\prime})-2\sum_{lm}%
e^{i\eta_{l}}\sin\eta_{l}Y_{l}(\hat{k})\cdot Y_{l}(\hat{k}^{\prime})\right]  ,
\end{equation}
where $\sigma=\pm1$ and $n(\varepsilon_{k\sigma})$ is the DOS for one
direction of the spin. Note that interchanging $\hat{k}$ and $\hat{k}^{\prime
}$ in the first term in the bracket changes its sign; this is the signature of
the antisymmetric scattering. The second term is the usual symmetric term
associated with charge scattering.

From the antisymmetric part of the scattering probability $W_{antisym}%
(k\sigma\rightarrow k^{\prime}\sigma)$, associated with cross terms between
the antisymmetric and symmetric parts of the $T$-matrix, we define
$\omega_{skew}(k_{F}\sigma)$ by,%

\begin{equation}
\sum_{k^{\prime}}W_{antisym}(\mathbf{k}\sigma\rightarrow\mathbf{k}^{\prime
}\sigma)g(\mathbf{k}^{\prime}\mathbf{,\sigma)\equiv}e\omega_{skew}(k_{F}%
\sigma)\mathbf{\hat{e}}\cdot\mathbf{\hat{k}\times\hat{z},}%
\end{equation}
where $\mathbf{\hat{e}}$ is a unit vectors along the electric field
$\mathbf{E}$, $\mathbf{\hat{z}}$ the spin quantization axis, $\mathbf{\tau
}_{0}$ the isotropic relaxation time and we have used the normal
out-of-equilibrium distribution function $g(\mathbf{k}^{\prime}\mathbf{,\sigma
)\equiv-}e\mathbf{\tau}_{0}v_{F}\mathbf{\hat{e}\cdot\hat{k}}^{\prime}$ to
arrive at this result. From the cross terms in the $T-$matrix between the
$l=2$ and $l\pm1$ terms and upon performing the integrals over spherical
conduction bands we find%

\begin{equation}
\omega_{skew}(k_{F}\sigma)=-\sigma\frac{6N_{i}\tau_{0}v_{F}}{\pi\hbar
n(\varepsilon_{F})}\frac{\lambda_{d}}{\Delta}\sin(2\eta_{2}-\eta_{1})\sin
^{2}\eta_{2}\sin\eta_{1},
\end{equation}
where $N_{i}$is the number of impurities.

The side jump contribution to the SHE enters when we write Hall current in the
presence of spin-orbit scattering~\cite{16} as $\mathbf{J}=-e\sum
_{\mathbf{k},\sigma}[\mathbf{v}_{\mathbf{k}}+\mathbf{\omega}_{a}%
(\mathbf{k},\sigma]f(\mathbf{k,\sigma)}$ where the term $\mathbf{\omega}_{a}$
is the anomalous velocity [arising from a side jump] attendant to electron
flow in systems with spin-orbit coupling \cite{17}. By using the distribution
function $f(\mathbf{k,\sigma)}$ found from the linearized Boltzmann equation
that accounts for antisymmetric (see Eq.~2) as well as symmetric scattering
and the appearance of side jumps~\cite{18}, we find the current can be written as%

\begin{align}
\mathbf{J}(\sigma &  =\uparrow\downarrow)=e^{2}E\sum_{\mathbf{k}}\left(
-\frac{\partial f^{0}}{\partial\varepsilon_{k\sigma}}\right)  \mathbf{\tau
}_{0}(k_{F}\sigma)[v_{F}\mathbf{\hat{k}}-\mathbf{\omega}_{a}(k\mathbf{,\sigma
)\hat{z}\times\hat{k})}]\\
&  \times\mathbf{\hat{e}\cdot\lbrack}v_{F}\mathbf{\hat{k}}+\{\mathbf{\omega
_{skew}(}k_{F},\mathbf{\sigma)}+\mathbf{\omega}_{a}(k_{F}\mathbf{,\sigma
)\}(\hat{z}\times\hat{k}))}],\nonumber
\end{align}
where we have written $\mathbf{\omega}_{a}(\mathbf{k},\sigma)=\mathbf{\omega
}_{a}(\mathbf{k},\sigma)\hat{k}\times\hat{z}$ .

The \textit{transverse} Hall current comes from terms proportional to
$\mathbf{\hat{z}\times\hat{e}.}$ When we consider spherical conduction bands
and average over $\Omega_{k}$ we find there are two transverse components in
$\sigma_{H}\equiv\sigma_{yx}$. The skew scattering one is%

\begin{equation}
\sigma_{skew}(\sigma=\uparrow\downarrow)=-\frac{1}{3}e^{2}\left[  \int
n(\varepsilon_{k\sigma})d\varepsilon_{k\sigma}\left(  -\frac{\partial f^{0}%
}{\partial\varepsilon_{k\sigma}}\right)  v(\varepsilon_{k\sigma})\mathbf{\tau
}_{0}(k_{F}\sigma)\mathbf{\omega_{skew}(}k_{F},\mathbf{\sigma)}\right]  ,
\end{equation}
and the anomalous velocity or side jump contribution is,%
\begin{equation}
\sigma_{anom}(\sigma=\uparrow\downarrow)=-\frac{2}{3}e^{2}\left[  \int
n(\varepsilon_{k\sigma})d\varepsilon_{k\sigma}\left(  -\frac{\partial f^{0}%
}{\partial\varepsilon_{k\sigma}}\right)  v(\varepsilon_{k\sigma})\mathbf{\tau
}_{0}(k_{F}\sigma)\mathbf{\omega_{a}(}k_{F},\mathbf{\sigma)}\right]  .
\end{equation}
The normal conductivity \textit{for each spin channel} is%

\begin{equation}
\sigma_{N}^{-1}=\frac{20\pi\hbar N_{i}}{n_{\sigma}e^{2}k_{F}}\sin^{2}\eta_{2},
\end{equation}
where $k_{F}$ is the momentum at the Femi level, and $n_{\sigma}=%
%TCIMACRO{\U{bd}}%
%BeginExpansion
\frac12
%EndExpansion
n_{total}$. By placing the expression for $\omega_{skew}(k_{F}\sigma)$ , see
Eq.~3, in the expression for $\sigma_{skew}$ , Eq.~5, and dividing by
$\sigma_{N}^{2}$ we find to first order in $\sigma_{H}/\sigma_{N}$ the skew
scattering contribution to the Hall effect is,%

\begin{equation}
\rho_{xy}^{skew}(\sigma=\uparrow\downarrow\rightarrow\pm)=\pm\frac{12\pi
N_{i}\hbar}{n_{\sigma}e^{2}k_{F}}\frac{\lambda_{d}}{\Delta}\sin(2\eta_{2}%
-\eta_{1})\sin^{2}\eta_{2}\sin\eta_{1}.
\end{equation}
The Hall angle from the skew scattering is%
\begin{equation}
\Phi_{H}^{skew}=\pm3/5\frac{\lambda_{d}}{\Delta}\sin(2\eta_{2}-\eta_{1}%
)\sin\eta_{1}.
\end{equation}

\begin{figure}[ptb]
\includegraphics[height=6cm]{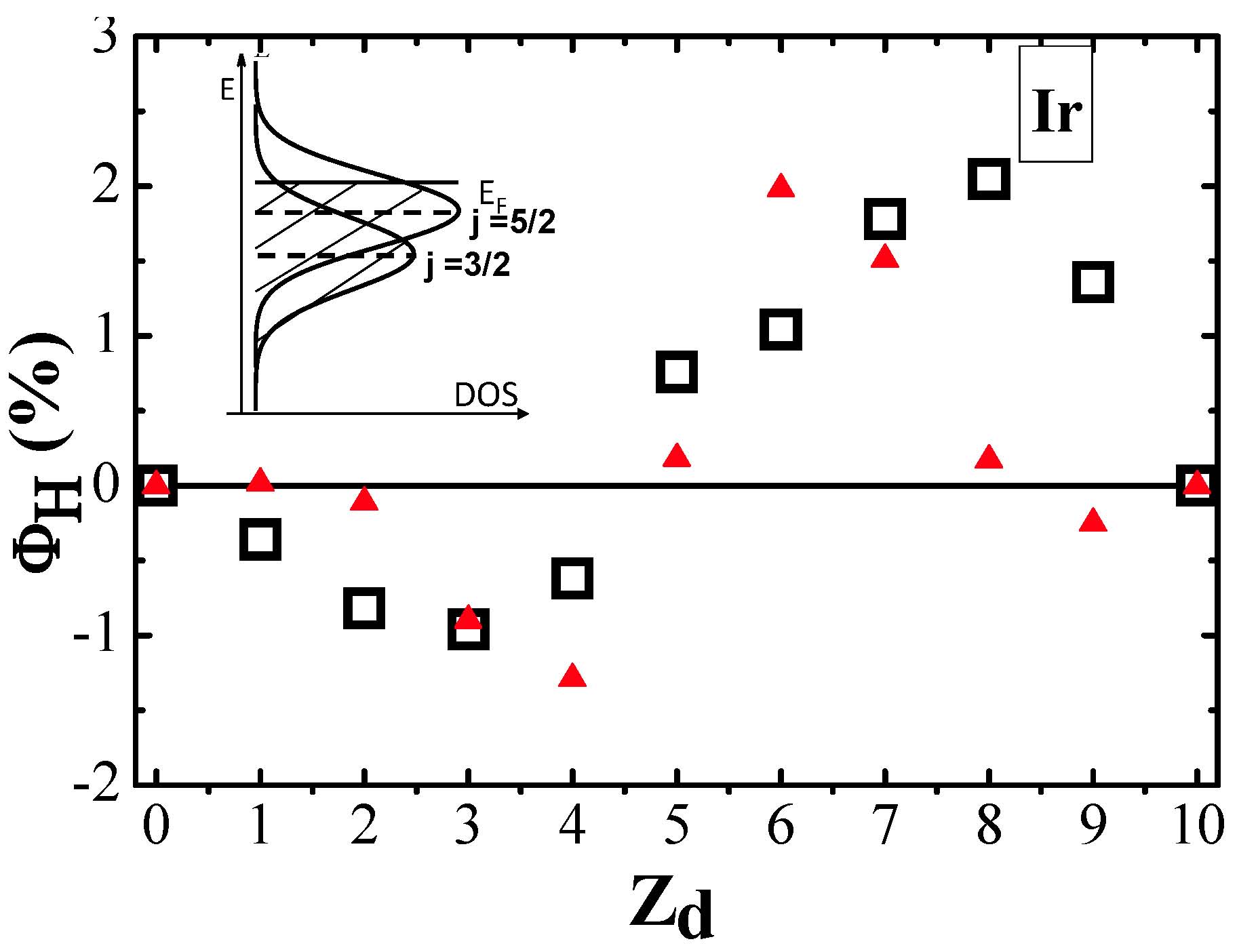}\caption{Skew scattering (squares) and side-jump (triangles) contributions to the Spin Hall Angle calculated from Eqs.9 and 13 as a function of the number of d electrons, $Z_{d}$, for $5d$ impurities in $Cu$ . The parameters are indicated in the text. The side-jump contribution is calculated for an impurity concentration of $2\%$ . Inset: Density of States
(DOS) of a $5d$ virtual bound state with S-O splitting between $j=3/2$ and
$j=5/2$ states.}%
\end{figure}

The contribution to the Hall effect from the side jump mechanism, i.e., the
anomalous velocity, is found by repeating the calculation done for the
Kondo-like rare-earth ions \cite{16}, but this time using the $T$ matrix for
resonant $d$ states, Eq.1. The expression for the anomalous velocity
$\mathbf{\omega}_{a}(\mathbf{k},\sigma),$Eq. (2.13) in Ref.16 is,%

\begin{align}
\mathbf{\omega}_{a}(\mathbf{k},\sigma)  &  =\frac{2N_{i}}{\hbar}%
[\operatorname{Re}\nabla_{\mathbf{k}}T_{\mathbf{k}\sigma,\mathbf{k}\sigma
}+\sum_{\mathbf{k\prime\sigma}^{\prime}}\mathit{P}\frac{1}{(\varepsilon
_{k\sigma}-\varepsilon_{k^{\prime}\sigma^{\prime}})}\operatorname{Re}%
T_{\mathbf{k}\sigma,\mathbf{k}^{\prime}\sigma^{\prime}}^{\dag}\nabla
_{\mathbf{k}^{\prime}}T_{\mathbf{k}^{\prime}\sigma^{\prime},\mathbf{k}\sigma
}\nonumber\\
&  -\pi\sum_{\mathbf{k\prime\sigma}^{\prime}}\delta(\varepsilon_{k\sigma
}-\varepsilon_{k^{\prime}\sigma^{\prime}})\mathit{\operatorname{Im}%
}T_{\mathbf{k}\sigma,\mathbf{k}^{\prime}\sigma^{\prime}}^{\dag}\nabla
_{\mathbf{k}^{\prime}}T_{\mathbf{k}^{\prime}\sigma^{\prime},\mathbf{k}\sigma
}].
\end{align}
Only the last term contributes to the Hall effect \cite{16} and by using
Eq.~1, we find%

\begin{equation}
\mathbf{\omega}_{a}(\mathbf{k},\sigma)=\sigma\frac{12N_{i}}{\pi n(\varepsilon
_{F})\hbar k_{F}}\frac{\lambda_{d}}{\Delta}\frac{E_{F}}{\Delta}\cos(3\eta
_{2}-\eta_{1})\sin^{3}\eta_{2}\sin\eta_{1}\mathbf{\hat{k}\times\hat{z}.}%
\end{equation}
By placing this expression in Eq.~6, and dividing by $\sigma_{N}^{2}$ we find
the anomalous velocity contribution to the Hall effect is,%

\begin{equation}
\rho_{xy}^{anom}(\sigma=\uparrow\downarrow\rightarrow\pm)=\mp\frac
{320N_{i}\hbar}{n_{\sigma}e^{2}k_{F}}\frac{c}{z}\frac{\lambda_{d}}{\Delta
}\frac{E_{F}}{\Delta}\cos(3\eta_{2}-\eta_{1})\sin^{5}\eta_{2}\sin\eta_{1},
\end{equation}
where $c$ is the impurity concentration, and $z\equiv\frac{n_{total}}{N_{s}%
}=\frac{2n_{\sigma}}{N_{s}}$ , i.e., the number of conduction electrons per
lattice site. Finally, the Hall angle from the side jump is,%

\begin{equation}
\Phi_{H}^{anom}=\mp16/\pi\frac{c}{z}\frac{\lambda_{d}}{\Delta}\frac{E_{F}%
}{\Delta}\cos(3\eta_{2}-\eta_{1})\sin^{3}\eta_{2}\sin\eta_{1}.
\end{equation}

Similar calculations can be performed in the presence of crystal field. With
completely crystal field split $t_{2g}$ and $e_{g}$ states, for example, the
prefactors of Eqs.~(9) and (13) for the $t_{2g}$ states are multiplied by
$\frac{1}{3}$ and $\frac{1}{5}$ respectively, and $\eta_{2}$ is replaced by
$\eta_{t_{2g}}=\frac{\pi}{6}Z_{t_{2g}}$ , where $Z_{t_{2g}}$is the number of
electrons in the $t_{2g}$ states.

We begin the discussion of our results by a glance at the expressions of the
Spin Hall Angle (SHA) for skew scattering and side-jump, respectively Eqs.~(9)
and (13), for the case without crystal field splittings. $\Phi_{H}^{skew}$ is
proportional to $\frac{\lambda_{d}}{\Delta}$, and $\Phi_{H}^{anom}%
\thicksim\frac{\lambda_{d}E_{F}}{\Delta^{2}}$. Large effects are thus expected
for narrow resonances when the S-O splitting induces significant differences
in the scattering on 5/2 and 3/2 states. In the corresponding expressions for
the intrinsic contribution to the SHA \cite{2,3} the denominator $\Delta$ is
replaced by an energy of the order of the band width, therefore extrinsic
effects due to resonant scattering should be generally larger in the usual
case where the width of the resonance is smaller than the band width.

The second important feature in the expressions for the Hall angle, arising
from the symmetry rules for the SHE, is the interplay between the asymmetric
scattering amplitude in the channel $l$ and the symmetric amplitudes in the
channels $l\pm\ 1$. It follows that the Spin Hall angle, Eqs.~(9) and (13),
depends not only on the phase shift $\eta_{2}$ in the resonant channel ($l$ =
2) but also on the phase shift $\eta_{1}$ in the non-resonant channel $l$ = 1
(we have neglected the phase shift in the channel with $l$ = 3). As the
scattering in a non-resonant channel is generally weaker than in a resonant
one, this selection of cross terms between different spherical harmonics
(rarely described in theoretical papers) contributes to the general smallness
of the SHA.

We now focus on the skew scattering. If one supposes, as generally admitted,
that the main contribution to the scattering by $5d$ impurities in noble
metals comes from the resonance on their 5d states~\cite{15}, $\eta_{2}$ is
much larger then $\eta_{1}$ and, in first approximation, $\Phi_{H}$ is
proportional to $\sin2\eta_{2}$; see Eq.~(9). As $\eta_{2}=\frac{\pi Z_{d}%
}{10}$, $\sin2\eta_{2}$ changes sign from positive to negative between the
beginning and end of the 5d series as shown in Fig.1. This change arises from
the difference in sign of the asymmetric resonant scattering on $5/2$ and
$3/2$ states. This agrees with the observed change of sign for the skew
scattering SHE induced by 5d impurities in Cu \cite{11,12}, $\Phi_{H}=-1.2\%$
for Lu impurities ($Z_{d}[Lu]=1$) and $\Phi_{H}=2.6\%$ for Ir impurities
($Z_{d}[Ir]=8$); the positive SHE for Ir in Cu has been confirmed by recent
experiments \cite{13}. A similar change in sign with $Z_{d}$ is observed also
for pure 5d metals, which suggests a similar explanation based on the relative
position of the $5d$ states, split by the S-O interaction, with respect to the
Fermi level. However,the crystal field splitting should also be taken into
account for a precise prediction of the variation in the $5d$ series. As we
have summarized after Eq.~13 the crystal field splitting between $t_{2g}$and
$e_{g}$ leads to a variation of the SHE as $\sin\frac{\pi Z_{t_{2g}}}{3}$ with
a change of sign for $Z_{t_{2g}}=3$.

Now, we proceed to a quantitative discussion of the skew scattering Hall angle
predicted by Eq.~(9). First we discuss $Cu$ doped with $Ir$ for which
different types of experiments have shown a predominant contribution from skew
scattering with reasonably consistent values of $\Phi_{H}$, $\Phi_{H}=2.6\%$
in Ref.\cite{11} and $\Phi_{H}=1.5\%$ in Ref.\cite{13} . Typical values of
$\Delta$ for $5d$ impurities in noble metals are close to $0.5eV$ from both
experiments~\cite{19} and ab-initio calculations \cite{9,20}. With
$\Delta=0.5eV$, $\lambda_{d}\approx0.25eV$ \cite{21}, and $Z_{d}=8$, the mean
experimental value of the SHA for $CuIr$, $\Phi_{H}=2.05\%$, is obtained by
introducing $\eta_{1}=-4.3%
%TCIMACRO{\U{b0}}%
%BeginExpansion
{{}^\circ}%
%EndExpansion
$ in Eq.~(9). The decomposition of Eq.~(9) into two factors, $\frac
{3\lambda_{d}sin(2\eta_{2}-\eta_{1})}{5\Delta}=-0.277$ and $sin\eta
_{1}=-0.075$, shows that the interference between the resonant and nonresonant
channels induces a significant reduction. The calculation for $Ir$ in $Cu$ can
be extended to other 5d impurities. With the same values of $\Delta$,
$\eta_{1}$, and using the S-0 constants $\lambda_{d}$ for the 5d series
\cite{21} and $\eta_{2}=\frac{\pi Z_{d}}{10}$, one obtains the wavy variation
of $\Phi_{H}$ as a function of $Z_{d}$ shown in Fig.1.

In contrast to the skew scattering contribution to the SHA, the side jump one
is proportional to the impurity concentration $c$. The side-jump SHA for
$c=2\%$, calculated with $E_{F}=7eV$ for Cu and the values of the parameters
$\lambda_{d}$, $\Delta$, $\eta_{1}$ already used for skew scattering, is
compared in Fig.1 with the skew scattering one. For impurities at the
beginning and the end of the $5d$ series ($Lu,Hf,Ir,Pt$) the side jump
contribution at $c=2\%$ is much smaller than the skew scattering one. It is
expected to remain smaller even at concentrations around $10\%$. This is in
agreement with the results of a constant SHA up to $c=12\%$ for $Ir$ in
$Cu$~\cite{13}. On the other hand, for impurities in the middle of the series,
like $W$,$Ta$ or $Os$, concentrations as small as $2\%$ yield side jump and
skew scattering contributions of the same order of magnitude. For these
impurities, very large contributions from the side-jump ($\Phi_{H}%
^{anom}\gtrsim10\%$ ) are expected for concentrations of the order of $10\%$.
This can be compared to the situation of the Anomalous Hall Effect of $Gd$
doped with $Lu$ impurities \cite{22} in which the side-jump contribution
exceeds the skew scattering for concentrations above about $6\%$ of $Lu$.

The above discussion, for both the skew scattering and side-jump
contributions, is altered when a crystal field splits the $t_{2g}$ and $e_{g}$
states. According to the results summarized after Eq.~(13), this introduces a
change of sign not in the middle of the $5d$ series but at midway through the
filling the $t_{2g}$ states at $Z_{t_{2g}}=3$ , and a reduction by 3 and 5 in
the amplitudes of the Hall angles. This can change the variation through the
$5d$ series but not really the order of magnitude of the SHA's. For
quantitative predictions only an $ab-initio$ calculation of the scattering
phase shifts can lead to realistic results. Our analytical calculation rather
aims to predict the main features of what can be expected from $5d$ resonances
and to identify the important parameters.

Summarizing, large SHE effects induced by the resonant scattering from
impurity states, here $d$ levels, are expected from the combination of: $i$) a
large S-0 coupling of the impurity states and a narrow resonance, which is the
condition to obtain a large asymmetric scattering amplitude in the resonance
channel $l$, and $ii$) a large symmetric scattering in the channels ($l\pm1$).
The second condition is not really fulfilled for $5d$ impurities in $Cu$ so
that skew scattering SHA's of only a few percent are expected in agreement
with the existing experimental results. However, at least for some impurities
($W,Ta,Os$), large side jump effects can be expected with $\Phi_{H}$ exceeding
$10\%$ for $c\gtrsim10\%$. We anticipate that similar features can also be
found for impurities with $p$ state resonances like $Pb$ or $Bi$. Spin Hall
angles above 10\% would be extremely interesting candidates for generating
spin currents without magnetic materials in spintronic devices.

We thank Professor Elie Belorizky for helpful discussions on the projection of
the scattering on crystal field states.

\bigskip


\begin{thebibliography}{99}
\bibitem {17}See Ref.16, Sec. \ IV. Note that while a magnetic field is used
in Ref.16 for the Hall effect, here we use a spin polarized current in which
$\hat{z}$ denotes the axis of spin polarization.
\end{thebibliography}
\end{document}